# State Compression and Quantitative Assessment Model for Assessing Security Risks in the Oil and Gas Transmission Systems

## Hisham A. Kholidy

**Department of Networks and Computer Security (NCS), College of Engineering, State University of New York (SUNY) Polytechnic Institute, United States of America. kholidh@sunypoly.edu.**

**Abstract:** The SCADA system is the foundation of the large-scale industrial control system. It is widely used in industries of petrochemistry, electric power, pipeline, etc. The natural gas SCADA system is among the critical infrastructure systems that have security issues related to trusted communications in transactions at the control system layer, and lack quantitative risk assessment and mitigation models. However, to guarantee the security of the Oil and Gas Transmission SCADA systems (OGTSS), there should be a holistic security system that considers the nature of these SCADA systems. In this paper, we augment our Security Awareness Framework (SAF) with two new contributions, (i) a Data Quantization and State Compression Approach (DQSCA) that improves the classification accuracy, speeds up the detection algorithm, and reduces the computational resource consumption. DQSCA reduces the size of processed data while preserving original key events and patterns within the datasets. (ii) A quantitative risk assessment model that carries out regular system information security evaluation and assessment on the SCADA system using a deductive process, where the topmost undesirable base attack event is postulated. Then, the ways for this event to occur are deduced. Our experiments denote that DQSCA has a low negative impact on the reduction of the detection accuracy (2.45% and 4.45%) while it reduces the detection time much (27.74% and 42.06%) for the Turnipseed's and Gao's datasets respectively. Furthermore, the mean absolute percentage error (MAPE) rate for the proposed risk assessment model is lower than the intrusion response system (Suricata) for the DOS, Response Injection, and Command Injection attacks by 59.80%, 73.72%, and 66.96% respectively.

**Keywords:** Cyber-attacks; oil and gas transmission SCADA; risk assessment; feature selection; data reduction.

## 1. Introduction

In the early period, the traditional SCADA system was a closed serial network that contained only trusted devices with little or no connection to the outside world. As control networks evolved, the use of TCP/IP and Ethernet became commonplace and interfacing with business systems became the norm. The result was that the closed trust model was no longer applied and vulnerabilities in these systems began to appear [1]. In particular, network security problems from the business. The natural gas and hazardous liquid pipelines are among these SCADA systems that are the core of the oil and gas field industrial control system, and its safety is related to the whole production and operation of the oil and gas field. The cyber infiltration of such SCADA systems could allow successful attackers to disrupt pipeline service and cause spills, explosions, or fires, all from remote locations. These intrusions have heightened congressional concern about cybersecurity in the U.S. pipeline sector. The natural gas and hazardous liquid pipelines SCADA has many cybersecurity issues in common with other critical infrastructures however, it is somewhat distinct in several ways [2]: (1) Pipelines have been the target of several confirmed terrorist plots and attempted physical attacks since September 11, 2001. (2) Changes to pipeline computer networks over the past 20 years, more sophisticated hackers, and the emergence of specialized malicious software have made pipeline SCADA operations increasingly vulnerable to cyber-attacks. (3) Recently, there has been a coordinated series of cyber intrusions specifically targeting U.S. pipeline computer systems. (4) The Transportation Security Administration (TSA) already has statutory authority to issue cybersecurity regulations for pipelines if the agency chooses to do so, but it may not have the resources to develop, implement, and enforce such regulations if they are mandated. Without an effective preventive measure installed in the OGTSS hostile foreign forces or hackers may launch cyber-attacks that destroy and cause irreparable consequences to these systems [3]. Due to the late emergence of information security problems in OGTSS and the wide range of evaluation involved, there is no systematic method to evaluate the information security at home and abroad. [4]

To the best of our knowledge, none of the current works introduces a holistic real-time security framework that specifically works for OGTSS networks and considers these systems' real-time scalability and dynamic features. None of the current approaches, particularly in the OGTSS security domain, quantitively assess the security risks and connects the attack-related events and system states with its risk assessment and mitigation processes, this is also called the situation awareness that should be considered at each risk assessment and mitigation point, nor considers the special OGTSS characteristics and requirements such as the criticality of the operation of the assets, high response impact and consequences, compliance requirements for regulatory agencies, and high level of scalability and interoperability that OGTSS maintains. To this end, this paper contributes towards, (1) developing a Data Quantization and State Compression Approach (DQSCA) that reduces the huge amount of data of the OGTSS dataset. This in turn improves the classification accuracy, speeds up the detection algorithm, and reduces the computational resource consumption. This approach reduces the size of the data while preserving original key events and patterns within the OGTSS dataset. (2) developing a quantitative risk assessment model that carries out regular system information security evaluation and assessment on the OGTSS using a deductive process, where the topmost undesirable attack base event is postulated. The evaluation of this model is based on the online attack scenarios that we implement in the testbed. These attack scenarios are described in Section 5.

Our new risk assessment model quantitatively assesses the risk in the SCADA and provides the required input parameters that will be used by the response system. It considers the aforementioned two purposes defined by the DHS. It measures the financial risk the SCADA assets face from cyberattacks by measuring the risk in terms of a numeric value that denotes the degree of security indices. Such a method can help assess how much security is improved if a specific response or security enhancement is applied. This in turns helps the response system to select between various enhancement choices, prioritize them by their relative effectiveness by measuring the improvement in the proposed degree-of-security indices, and make cost justifications. The proposed risk assessment model also helps set and track a specific numeric target for security level and uses a Hierarchical Risk Correlation Tree (HRCT) diagram as a graphical illustration showing the stepwise cause resolution using formal logic symbols. The evaluation of the tree is done quantitatively using a probabilistic evaluation which computes the probability of occurrence of the top events. The proposed model assesses the risk in the infrastructure based on the alert level of different events by measuring the potential impact of a threat on assets given the probability that it will occur, and it provides useful information to evaluate the system's overall security state. The estimated risk of each event is not assigned statically; rather it is assigned an initial value that is modified dynamically as the event is correlated to other ones. HRCTs are designed offline by experts on each computing asset, e.g., a PMU or a PDC, residing in a SCADA network. Unlike the current attack trees that are designed according to all possible attack scenarios, HRCT is built based on the attack consequences, e.g., a denial of service against PDC/PMU. Thus, the HRCT does not have to consider all possible attack scenarios that might cause those consequences. This in turns results in a fewer number of leaf nodes and consequently reduces the cardinality exponential growth of the system security state space that usually causes the state space explosion problem. Both DQSCA and risk assessment model help solving the state space explosion problem and enable the security framework to provide a timely decision.

The remainder of this paper is organized as follows. Section 2 presents a survey about the current risk assessment model and SCADA security solutions. Section 3 describes the OGTSS testbed. Section 4 discusses our existing security framework. Section 5 highlights the attack scenarios that are used to evaluate the proposed work. Section 6 introduces the DQSCA. Section 7 evaluates the performance of the DQSCA. Section 8 introduces the HRCT model. Section 9 introduces a case study that evaluates the HRCT model using the command injection attack scenario. Finally, Section 10 draws some concluding remarks and outlines future work.

## 2. LITERATURE REVIEW

The literature review [5, 6, 7, 8, 9, 10] revealed that there are two types of risk analysis methods, qualitative and quantitative methods. The former methods are often selected by managers who believe their risk-assessment calculations are simple; therefore, it is not necessary to quantify threat frequency, hence many nontechnical issues are easily accounted for. [9] The latter is used to evaluate the investments in security enhancements by considering the general cost-benefit analysis methods that compare the cost of deploying the security to the benefits [10]. However, there is a limited amount of work on the development of a quantitative risk assessment system that adaptively works for the SCADA systems and quantitatively assesses the financial damage of the attacks, and computes the severity of such attacks in these systems. [10] According to the Department of Homeland Security [11], a formal risk assessment of information security serves two purposes: (1) to identify existing weaknesses in the systems, and (2) to cost-justify and prioritize the cost of additional safeguards.

In [12], the authors analyzed twenty-four models and described research challenges while developing risk assessment methods for SCADA systems. According to [12], the assessment methods pay little attention to a profound understanding of SCADA systems, and the lack of accurate data while calculating probabilities results in inaccurate risk. Moreover, the SCADA system lacks data on cyber incidents for accurately define risk, and lack of elaboration and presentation of risk assessment methods using software tools shows the immaturity of risk assessment research in SCADA systems.

In [13], the authors investigated cyber threats targeting physical systems. They proposed a classification based on five parameters (types of attack, target sector, intention, impact, and incident category). They provided a matrix of these threats in conjunction with simple statistical data. In [14], the authors identified fifteen types of threads and four SCADA system components. First, they linked between each thread and target component, and then they determined the vulnerabilities for each system component based on historical data and the component's characteristics.

## 3. OGTSS TESTBED

Fig.1 shows the Cyber-Physical System testbed architecture. It has software and hardware components. The software component LabVIEW RT simulates the physical system, namely the Continuous Stirred Time Reactor (CSTR) in real-time. On the hardware side, the Basic Process Control System (BPCS) and Safety Instrumented System (SIS) uses an I/O physical data bus to exchange the reactor's simulation data.

Both the controllers run RT (real-time) Linux OS. The BPCS implements process control functions. Likewise, the SIS implements the process shutdown logic functions. A control network interconnects BPCS, SIS, Human Machine Interface (HMI) workstation, and the Engineering workstation. Modbus protocol is used to allow BPCS and HMI to control the physical processes. A firewall is configured to isolate the control network from the corporate network, and the historian and real-time data servers are kept in a Demilitarized Zone (DMZ).

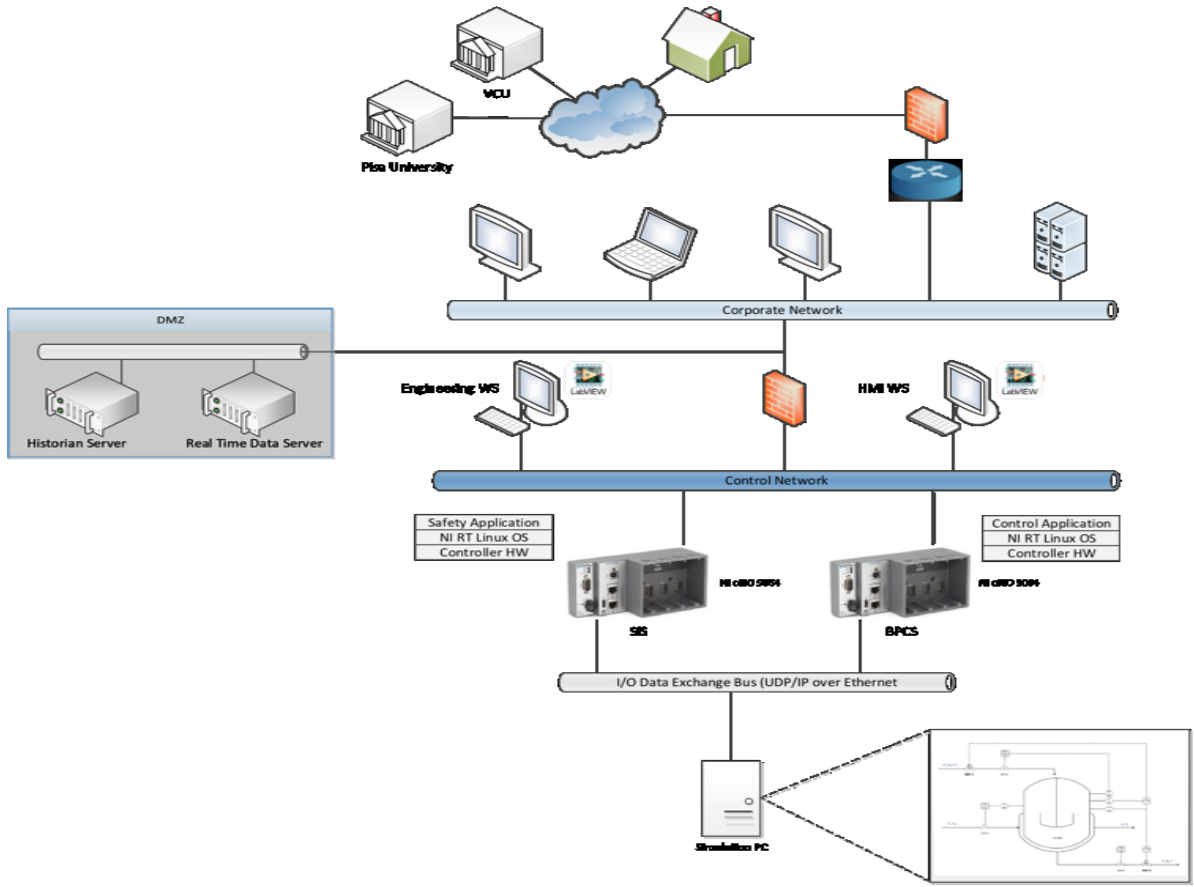

**Fig. 1**. OGTSS testbed Architecture

The Piping & Instrumentation Diagram (P&ID) for the reactor is depicted in Fig.2. The inlet or feed flow is controlled by either SDV-1(Shutdown Valve) or CV-1 (Control Valve) or both simultaneously. Likewise, CV-2 opens or closes for the coolant flow, and SDV-2 and CV-3 control the outlet flow. Shutdown valves detect the reactor level and act accordingly. For instance, if the level exceeds the maximum limit, then SDV-1 stops the inlet flow.

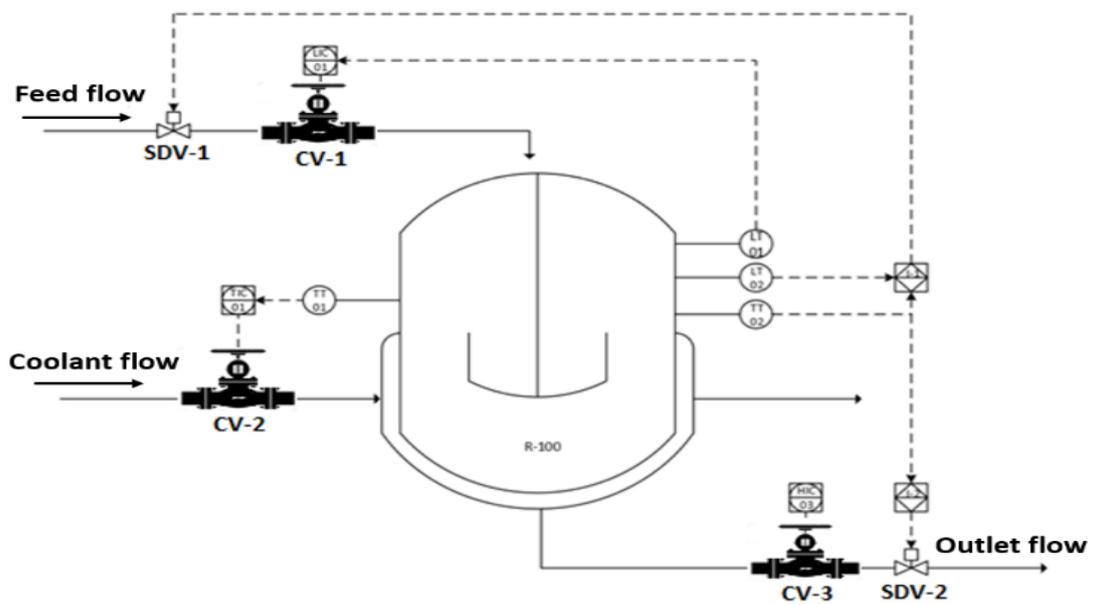

**Fig. 2.** Reactor P&ID

## 4. OUR EXISTING SECURITY FRAMEWORK

In the following, we give a high-level description of our security framework components and processes, see Fig. 3. In this paper, we will focus only on the risk assessment and data reduction processes. The full details of the security framework are given in [15-31].

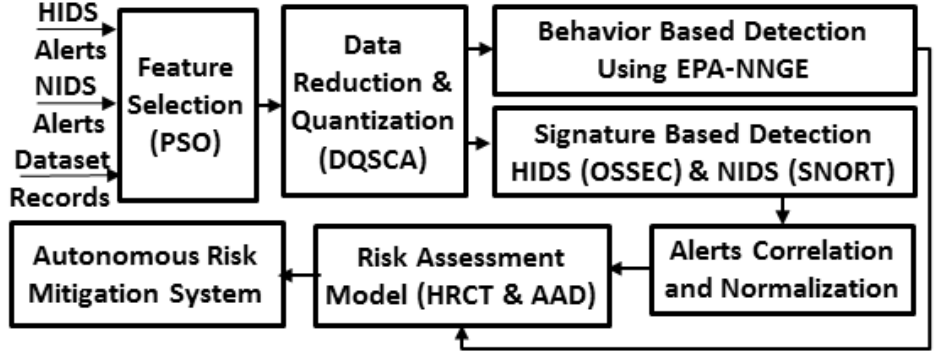

**Fig. 3.** Our security framework components and processes

### A. COLLECTION.

This process collects events and logs from several signatures-based sensors and sends them to the integration process. The collection sensors perform three core functions through various means: collecting logs, monitoring network packets, and scanning hosts. In our testbed, there is a NIDS sensor that extends Snort to monitor the network traffic flowing through the corporate and the control network switch. This process produces snort logs. Snort continually monitor the Modbus commands and responses which are controlling the physical processes. We have updated the snort with Modbus rules.

### B. INTEGRATION.

This process integrates the events collected from distinct signature-based sensors through two processes, namely, normalization and prioritization. The former formats any sensor event into the IDMEF protocol format [32] to facilitate the analysis and correlation of these events in the next layer. The latter handles the prioritization systems of distinct detectors i.e., host and network IDSs. This process maps the alert severity (i.e., how dangerous the alert is) of each detector into one common range from $0$ to $n$, where n is defined by system administrators.

### C. CORRELATION

It correlates the normalized events from different sensors to highlight the few critical ones. It compares each event against a set of attacks' rules to discover if it signals a true attack and then it correlates the related events. Events are related if they have the same source and destination addresses are close in time and denote the same attack signature.

### D. FEATURE SELECTION

This process extracts a subset of relevant features from the relay and controller logs to enhance the classification results. In [33], we introduced an approach that accurately selects the most relevant features and ignores the irrelevant ones from the input log files using the Particle Swarm Optimization (PSO) approach. This approach helps in the elimination of the possibility of incorrect training through the removal of redundant features and noises.

### E. BEHAVIOR BASED DETECTION

We have introduced a new behaviour-based detection approach that combines the PSO with the Evolutionary Pruning Algorithms (EPA-NNGE) [8, 34] to improve the classification accuracy, speed, and reduces the computational resource consumption of the NNGE algorithm through pruning of the non-generalized exemplars using the highest-ranked features of the PSO. The behaviour-based detec-tion approach reduces the model size by reducing the state space cardinality exponential in the number of the attributes used to describe the protected system by reducing the hyper-rectangles and ignoring the non-selected features among the selected significant ones defined by the new fitness function of the PSO. For more details about the detection approach, see [7, 8].

*F. RISK ASSESSMENT*

The framework assesses the risk in the infrastructure based on the alert level of different events. The proposed model measures the potential impact of a threat on assets given the probability that it will occur, and it provides useful information to evaluate the system's overall security state. Section 8 describes the proposed risk assessment model using the HRCT in detail.

*G. Autonomous Risk Mitigation System.*

This process selects the most suitable set of countermeasures to protect the hosts and the network against an attack. The decision making considers the current system risk state computed by the HRCT, the criticality of the attack, the benefits of the countermeasure plan in terms of the OGTSS asset value that can be protected by applying the countermeasure plan, and the cost of the OGTSS asset that can be negatively affected when applying that countermeasure plan. The autonomous risk mitigation system is out of the scope of this paper.

## 5. ATTACK SCENARIOS

We used the OGTSS testbed to simulate attacks from the following four categories.

*A. RECONNAISSANCE ATTACKS.*

Passive attacks are essential to gather all operational information about the testbed. An attacker must be sure about the protocols and their parameters before constructing any active attack. Several tools are available to perform passive attacks. The most popular one is Metasploit, which has modules related to SCADA networks. After running multiple passive attacks, we found that the testbed is using Modbus TCP protocol and MTU (Master terminal unit) and RTU (Remote Terminal Unit) are using unit-id '1'.

*B. SPOOFING ATTACK.*

This attack enables an attacker to deceive a system, to believe that an attacker is a legitimate entity. In a SCADA system, an attacker pretends to be an MTU or RTU. As a result, that attacker able to exchange and falsify the network traffic between all the nodes. In our experiment, we have used the ARP poisoning attack against the MTU and RTU. Now, all the traffic is going through the attacker's machine.

*C. SNIFFING ATTACK.*

Such attacks allow an attacker to capture network packets. Network packets contain information about a network such as protocols, IP addresses, ports, and many more. In our testbed, after running a spoofing attack, MTU and RTU send or receive packets to or from the attacker's machine which now acts as a man in the middle.

*D. DOS ATTACKS.*

DoS (Denial of Service) attacks on the SCADA system cause catastrophic damages. In the industrial control system, the availability of resources is the primary goal. DoS attacks make the system resources unavailable. In our testbed, after performing the sniffing attack the attacker strategized the DoS attacks by analysing the gathered information. It is seen from the sniffing attack that the MTU and RTU use Modbus over TCP. So, to completely disrupt the link between them, an attacker must attack the TCP connection. Likewise, for disrupting only control operation, an attacker must attack the Modbus packet exchange. We have designed an attack on TCP. It stops the TCP connection between MTU and RTU and stops all the processes temporarily. When we retracted the attack, all came to normal. Similarly, we were able to stop the control operation by attacking Modbus.

*E. COMMAND INJECTION ATTACKS.*

A malicious command injection attack can change the state of a control valve. Three control valves are there to control the reactor level and temperature. To cause any process hazard, it needs to design active attacks to open/close the actuators (control valves). In a sniffing attack, we analysed the commands and responses to design malicious command injection attacks. By using Metasploit modules, we then sent these commands to the BPCS controller to open or close the control valves. As a result, the reactor level and temperature increase or decrease according to the position of control valves.

*F. RESPONSE INJECTION ATTACKS*

In the Modbus protocol, RTUs send back responses to MTU. According to the commands from MTU, RTU responses. In the testbed, Modbus used for communication between RTU and MTU. The Modbus protocol has some loopholes such as a lack of authentication that may allow a malicious entity to act as a man in the middle and send false responses to MTU. In building a response injection attack, we first diverted the traffic between MTU and RTU using a spoofing attack and then route the traffic through an attacker machine. We designed malicious responses from the control logic we learned. Finally, we keep inserting these malicious responses to interrupt the control processes.

## 6. DATA REDUCTION and QUANTIZATION

The new DQSCA horizontally reduces the size of the processing data (dataset records) resulted from the feature selection process while preserving original key events and patterns within the datasets using a DQSCA is a new data processing and compression method to quantize and compress the heterogeneous datasets while preserving original key events and patterns within the datasets. DQSCA tracks system states from measurements and creates a compressed sequence of states for each observed scenario. It pre-processes the OGTSS events to minimize the state space and this, in turn, reduces the number of rules generated by EPA-NNGE. This results in high detection accuracy, a short detection time, and a short time to build the model.

DQSCA receives as input, raw data with both the discrete and continuous data format. This input may come from separate datasets with a set of input features or may come in the form of separate streams of log events that are collected across multiple sensors, see Fig. 3. DQSCA converts the reduced datasets resulted from the feature selection process into one list that has a continuous stream of ordered states and labeled events.

**Step 1: Collecting Raw Data.** This data consists of time stamps and measurements. Equations 1, 2, and 3 show three timestamps and measurements from two example OGTSS sensors, S1 and S2. E.g., $S1_1$ means at timestamp 1, measurements are received from sensor S1. $s2_{a1.5}$ denotes at timestamp 1.5, a measurement from sensor s2 is received for item 'a'.

$$S1_1 = (s1_{a1}, s1_{b1},\ldots\ldots,t_{s11}) \quad (1)$$

$$S2_{1.5}= (s2_{a1.5}, s2_{b1.5},\ldots\ldots,t_{s21.5}) \quad (2)$$

$$S1_2=(s1_{a2}, s1_{b2},\ldots\ldots,t_{s12}) \quad (3)$$

**Step 2: Merging Raw Data.** One database will store all merged sensor data. The merged data must be time aligned because the measurements that each sensor takes are at different times. A baseline is defined according to the sensor with the highest frequency. In this way, the baseline sensor's log file will store the slower rate sensor data that will be merged.

**Step 3: Quantizing data.** OGTSS datasets have heterogeneous data from variant sensors. This data can take many forms; integers, Boolean, real, etc. to reduce state space, data must be quantized. Sensors that produce integer and real values data, the quantization will be based on numbered ranges. E.g., current and voltage can be quantized into *high (2), medium (1), and low (0),* and are ranged based on two thresholds $thr_1$ and $thr_2$. Expert knowledge is required to choose these thresholds. E.g., a quantization mapping for measurements $Q(s_i)=0$ if $s_i \leq thr_1$, $Q(s_i)=1$ if $thr_1 < s_i < thr_2$, and $Q(s_i)=2$ if $s_i \geq thr_2$.

**Step 4: Mapping to states.** A state is a set of quantized and merged sensor measurements and a time stamp. E.g., $S_j=(Q(s1_i), Q(s2_i),…t_i)$. All unique states are stored in a state database that is common for all instances of all scenarios. The index of the state, $j$, is incremented for each inserted unique state. After applying the mapping process, each instance is referred to as an un-compressed list of states. E.g., $U_n = (S_1, S_2, S_2, S_4,…)$ is an un-compressed list of states that represents the $N^{th}$ instance of scenario U.

**Step 5: Compressing data into state lists. DQSCA** compresses the un-compressed list by removing any sequence of states that do not change and just leave only one instance of that state. Furthermore, all rows in the resulted dataset with no change from the previous rows are removed. We use a time window to compare the repetition of the states between the rows. This in turns reduces memory usage and results in a sequence path that represents all state transitions for the system that correspond to the system events. E.g., a path $P_i$ is a list of observed system states $<S_0, S_1, S_2… S_n>$ that are arranged according to their timestamps. For a dynamic system such as OGTSS, DQSCA will produce many paths for every single scenario due to minor variations in sampled data which usually resulting from inaccurate measurements and contiguous changes in such systems. Consequently, DQSCA should periodically run to update the system state database and the quantization thresholds.

## 7. Evaluating the DQSCA Performance.

### A. The OGTSS Datasets

To evaluate the DQSCA, we use two datasets, Turnipseed's dataset [35] and Gao's dataset [36], that support the IDS research for SCADA systems. The former dataset contains three separate categories of features: network information, payload information, and labels. The network information provides a pattern of communication for intrusion detection systems to train against. The payload information provides information about the gas pipeline's state, settings, and parameters. The network transaction label is appended to each line in the dataset to detail if the transaction is normal system activity or an attack. The Turnipseed's dataset consists of two sub-datasets that include network traffic captured on a gas pipeline SCADA system and contain a total of 274,627 instances in each dataset. Each row in the dataset contains multiple columns, which are commonly referred to as features. The first sub dataset is a raw unprocessed dataset that contains raw network traffic data. There are six features for each instance in the raw dataset. The first feature contains the Modbus frame that was received by either the master or slave device. The Modbus frame contains all information from the network, state, and parameters of the gas pipeline. (Master/Slave address, Function Code, Register Address, Value to write, CRC). The second and third features in a raw dataset row represent the category of attack and specific attack that was executed. The fourth and fifth features in a raw dataset row represent the source and destination of the frame. The last feature in the raw dataset contains a time stamp. Fig.4 shows an example of a row from the raw dataset.

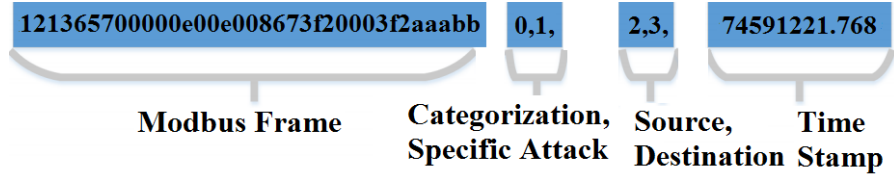

**Fig. 4**: Example for an instance row within Raw Dataset

The second sub-dataset is the Attribute Relationship File Format (ARFF) dataset. It contains twenty features, some of which are the same as in the raw dataset. These features are, the first feature contains the station address of the slave device, the second feature contains the function code. E.g., read command (0x03) and write command (0x16). The third feature contains the Modbus frame length. The fourth feature contains the setpoint value that controls the pressure in the gas pipeline. The next five features represent the PID controller values. Gain, reset rate, dead band, cycle time, and rate are all values that are used to tune the PID controller. The tenth feature contains the value which controls the system's duty cycle (i.e., 0 for Off, 1 for Manual, and 2 for Automatic). The twelfth feature controls the pump state only if the system mode is set to manual. The feature can only be two values off, '0', or on, '1'. The twelfth feature controls the pump state only if the system mode is set to manual. The feature can only be two values off, '0', or on, '1'. The thirteenth feature controls the state of the solenoid valve when the system is also in manual mode. There are only two possibilities for this feature '0', closed, and '1', opened. The fourteenth feature contains the current pressure measurement from the gas pipeline. The fifteenth feature contains the cyclic redundancy check (CRC) that allows the system to check for errors within a frame that is being provided to either the master or the slave device. The sixteenth feature is an indicator flag that helps the IDS to learn the difference between commands and responses. The value can either be a '0' for a response or '1' for command. The last four features, timestamp, specific attack, category attack, and binary attack, were also provided in the raw dataset.

The latter dataset, Gao's dataset, includes network transactions between a Remote Terminal Unit (RTU) and a Master Control Unit (MTU) in Mississippi State University's in-house SCADA gas pipeline. Gao's dataset includes four data sets. In our experiments, we will use only dataset I that contains transactions from the gas pipeline system.

Both Turnipseed's dataset and Gao's dataset have the same seven attacks that are split into four overall categories: command injection, response injection, denial of service (DoS), and reconnaissance. The categories of attacks contained in the dataset are shown in Table 1.

Table1: The seven attacks in the OGTSS dataset and their overall categories

| Attacks | Category |
| --- | --- |
| Naïve Malicious Response Injection. | Response Injection |
| Complex Malicious Response Injection. | Response Injection |
| Malicious State Command Injection. | Command Injection |
| Malicious Parameter Command Injection. | Command Injection |
| Malicious Function Code Injection | Command Injection |
| Denial of Service | Denial of Service |
| Interruption Reconnaissance | Reconnaissance |

*B. Experiment Results*

In this section, we evaluate the impact of reducing the dataset instances using the DQSCA on the classification accuracy, model size, and computational performance of the EPA-NNGE algorithm. The first step in our experiments is to use the modified PSO algorithm with the EPA-NNGE, as described in [7, 8], to select the most significant features in the Turnipseed's and Gao's Datasets.
Table 2 lists these features and their corresponding PSO fitness values as following, 11 features from Turnipseed's dataset (out of 20 features) and 15 features from Gao's dataset (out of 27 features), see Table 2. The modified PSO algorithm has almost the same reduction ratio (55%) for the two datasets.

After selecting the important features, the following DQSCA steps are applied to quantize and compress the datasets:

1) We use the significant features selected in the previous experiment. The numeric values of each feature are then quantized using domain expert inputs. Table 3 demonstrates some examples of the quantization intervals.
2) Assigning states to each sample as following:
    a. Any row of the quantized data that contains the same quantization values for each feature is assigned the same state.
    b. The states of each data are inserted into the quantized datasets using the selected features.
    c. The duplicated values (i.e., the records in the dataset have the same state and marker values) are deleted. The unique records are only saved. Using this step, the size of the dataset was reduced from 5MB raw datasets to 6KB. This reduction can significantly enhance the detection speed.

TABLE 2: The Most Significant Selected Features Fitness Values in the Turnipseed's and Gao's Datasets.

| Turnipseed's Dataset | | | Gao's Dataset | | | | | |
|---|---|---|---|---|---|---|---|---|
| Order | Feature name | Fitness value | Order | Feature name | Fitness value | Order | Feature name | Fitness value |
| 1 | Command response | 18.9 | 1 | Measurement | 22.9 | 12 | response_memory | 14.6 |
| 2 | Pressure measurement | 17.7 | 2 | Control_mode | 21.7 | 13 | Solenoid | 13.9 |
| 3 | Length | 17.2 | 3 | Resp_read_fun | 20.7 | 14 | resp_length | 11.3 |
| 4 | Setpoint | 16.4 | 4 | Control_scheme | 19.4 | 15 | resp_write_fun | 11.3 |
| 5 | Gain | 15.1 | 5 | Pump | 19.1 | | | |
| 6 | Reset rate | 15.1 | 6 | Setpoint | 17.3 | | | |
| 7 | Deadband | 15.1 | 7 | Comm_read_functio | 16.8 | | | |
| 8 | Cycle time | 15.4 | 8 | Command_memory_count | 15.4 | | | |
| 9 | System mode | 14.5 | 9 | Command_memory | 15.4 | | | |
| 10 | Control scheme | 13.4 | 10 | Command_address | 15.1 | | | |
| 11 | Pump | 12.3 | 11 | response_memory_count | 14.6 | | | |

DQSCA minimizes the memory requirements and processing time when validating classification algorithms. This enables the risk assessment model to provide a timely quick risk assessment as desired. DQSCA reduces the Turnipseed's dataset from a total of 274,627 instances to only 179328 instances (34.7%).

TABLE.3: Example for Measurement Quantization in the Turnipseed's and Gao's Datasets.

| Turnipseed's Dataset | | | | Gao's Dataset | | | |
|---|---|---|---|---|---|---|---|
| Feature | Quantization Interval Name | Interval Enumeration | Range | Feature | Interval Name | Interval Enumeration | Range |
| Command/response | '0' for a response or '1' for a command | {0,1} | {0,1} | Measurement | Low, Normal, High | {0,1,2} | {[-3.71E+19 , 0.65E+35), [0.65E+35, 3.65E+35), [3.65E+35, 6.65E+35)} |
| Pressure measurement | Low, Normal, High | {0,1,2} | {[1418682163, 1418764100), [1418764100, 1418807854), [1418807854, 1418964100)} | Control_mode | Automatic, manual, or shutdown | {0,1,2} | {0,1, 2} |
| Length | Short, Medium, Large | {0,1,2} | {[10, 12, 14, 16), 46, 90} | Resp_read_fun | Value of response function | {1,3} | {1, 3} |
| Setpoint | Low, High | {0,1} | {[0, 20), [20, 50)} | Control_scheme | Compressor (activated/ deactivated) | {0,1} | {0,1} |
| Gain | Low, High | {0,1} | {[0, 67), [110, 119)} | Pump | Pump is on or Off | {0,1} | {0,1} |

Table 4 shows an example of the selected features of 4 instances of the Turnipseed's dataset that are quantized into one instance with the following enumerations "2, 0, 2, 0, 1, 0, 0,0,0, 2". The 4 instances fit with the same features interval names *(large, low, high, low, medium, short, 0, 0, 0, high)* as demonstrated in table 3. DQSCA also reduces Gao's dataset from a total of 97019 instances to only 70242 instances (27.59%, 31 duplicate instances are removed and 26746 are combined using the quantization feature).

Table. 4: An Example of the Quantized Instances of the Turnipseed's dataset.

| 90 | 14.5 | 111 | 0.27 | 0.45 | 1 | 0 | 0 | 0 | 1418802029 |
| 90 | 11.6 | 111 | 0.23 | 0.55 | 0.8 | 0 | 0 | 0 | 1418850058 |
| 90 | 13.98 | 111 | 0.33 | 0.60 | 0.87 | 0 | 0 | 0 | 1418932698 |
| 90 | 10.5 | 111 | 0.20 | 0.61 | 1.26 | 0 | 0 | 0 | 1418852258 |

Table 5 shows an example of 4 instances that are quantized into one instance "4,*CmdOutMem, ResInMem*, 9, 18, 3, 3, *Short*, 19, *low*, 0, 1, 0, 0, 1". The 4 instances have the same quantized measurement feature interval with the normal interval of enumeration 1 as demonstrated in table 3.

Table. 5: An Example of the Quantized Instances of the Turnipseed's dataset.

| 4 | 183 | 233 | 9 | 18 | 3 | 3 | 10 | 19 | 20 | 0 | 1 | 0 | 0 | 1.41E-45 |
| 4 | 183 | 233 | 9 | 18 | 3 | 3 | 10 | 19 | 20 | 0 | 1 | 0 | 0 | 1.12E-43 |
| 4 | 183 | 233 | 9 | 18 | 3 | 3 | 10 | 19 | 20 | 0 | 1 | 0 | 0 | 1.54E-44 |
| 4 | 183 | 233 | 9 | 18 | 3 | 3 | 10 | 19 | 20 | 0 | 1 | 0 | 0 | 1.68E-44 |

The feature selection approach reduces the detection time of the EPA-NNGE algorithm by 11.37% and 11.41% for both the Turnipseed's and Gao's datasets respectively. This detection time has further reduced using both the DQSCA and the feature selection approach by 39.11% and 53.47% for both the Turnipseed's and Gao's datasets respectively. For online detection, DQSCA periodically updates the quantization intervals, and the corresponding states of each data are also updated and inserted into the quantized datasets and duplicated values are deleted.

Table 6 compares the EPA-NNGE detection accuracy using the feature selection approach only and the EPA-NNGE detection accuracy with both the DQSCA and feature selection approaches. The EPA-NNGE accuracy rate for each dataset is computed as a ratio between the numbers of correctly classified records/instances to the total number of records/instances in each dataset. Figure 5 shows an overall view of the EPA-NNGE detection accuracy ratio, DQSCA time reduction ratio, and feature selection time reduction ratio for the Turnipseed's and Gao's datasets. The detection accuracy reduction rate using the feature selection approach and the DQSCA is computed as *Acc(EPA-NNGE (DQSCA,PSO) - Acc(EPA-NNGE(PSO))*. As shown in Table 6, the lowest accuracy reduction rate (2.45%) was obtained using Turnipseed's dataset. This can be explained by the fact that this dataset has the lowest instances reduction ratio using the DQSCA and almost the same feature reduction ratio using the PSO. This denotes that DQSCA reduces the detection accuracy a little bit (2.45% and 4.45%) while reduces the detection time much (27.74% and 42.06%) for the Turnipseed's and Gao's datasets respectively. We can also notice that the DQSCA reduces the Gao's datasets instances very much compared to Turnipseed's dataset because Gao's dataset has more extracted features and a larger number of duplicated records that DQSCA removed.

TABLE 6: A Comparison between the EPA-NNGE with the feature selection only and the EPA-NNGE with both DQSCA and the feature selection.

|  | Turnipseed's dataset | Gao's Dataset |
|---|---|---|
| **Accuracy Reduction Ratio (%)** | -2.45 | -4.54 |
| **EPA-NNGE Detection Rate (PSO+DQSCA)(%)** | 91.87 | 88.84 |
| **Detection Rate (PSO) (%)** | 94.32 | 92.38 |
| **Time Reduction Rate (%)** | 39.11 | 53.47 |
| **Detection Time Reduction Rate (PSO) (%)** | 11.37 | 11.41 |
| **Detection Time Reduction Rate (DQSCA) (%)** | 27.74 | 42.06 |

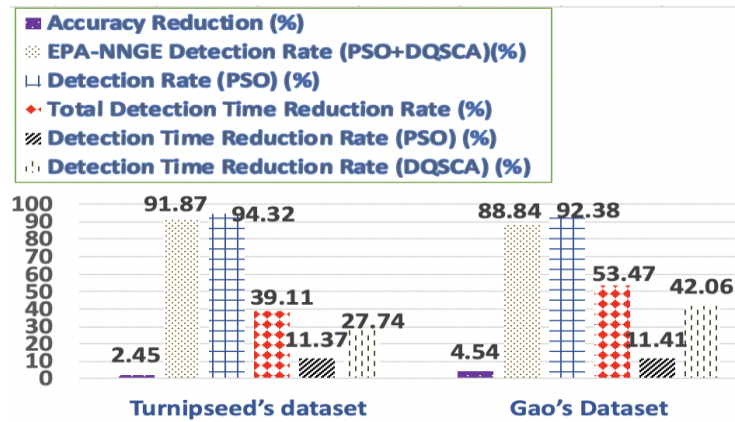

**Fig.5**: The EPA-NNGE detection accuracy ratio, DQSCA time reduction ratio, and feature selection time reduction ratio for the Turnipseed's and Gao's datasets.

# 8. The HIERARCHICAL RISK CORRELATION TREE (HRCT) MODEL

The proposed risk assessment model quantitatively assesses the risk in the SCADA and provides the required input parameters that will be used by the intrusion response system. It measures the financial risk the SCADA assets face from cyberattacks by measuring the risk in terms of a numeric value that denotes the degree of security indices. Such a method can help assess how much security is improved if a specific countermeasure or security enhancement is applied. The proposed risk model is a probabilistic risk assessment model that is built on the fact that complex or multi-stage attacks are a sequence, e.g. a chain, of elementary attacks where a threat agent acquires the privileges to implement each attack through the previous attacks in the chain such attacks are enabled by a global vulnerability, a set of local vulnerabilities that are correlated because the attacks they enable can be composed into a sequence. To build our risk model, we use a deductive process where the topmost undesirable base event is postulated. Then, the ways for this event to occur are deduced. The deduction process results in our Hierarchical Risk Correlation Tree (HRCT) that includes all components that could contribute to causing the top event. The HRCT is a graphical illustration showing the stepwise cause resolution using formal logic symbols. The HRCT models the paths an attacker can traverse to reach certain goals that adversely affect the OGTSS. Our motive in designing HRCT is to provide the intrusion response system with an accurate risk evaluation to proactively prevent an attacker from moving from one attack goal node to another, by responding appropriately at specific nodes. The estimated risk of each event in the HRCT tree is not assigned statically; rather it is assigned an initial value that is modified dynamically as the event is correlated to other ones. HRCTs are designed offline by experts on each computing asset, e.g., a PMU or a PDC, residing in a SCADA network. Unlike the current attack trees that are designed according to all possible attack scenarios, HRCT is built based on the severity of the attack consequences, e.g., a denial of service against PDC/PMU. Thus, the HRCT does not have to consider all possible attack scenarios that might cause those consequences, instead, it uses a severity-based policy (i.e., aggressive, moderate, and conservative policies) to select some potential attack events based on the reliability of the fired alert. This in turns results in a fewer number of leaf nodes in the tree and consequently reduces the cardinality exponential growth of the system security state space that usually causes the state space explosion problem.

In our experiments we set severity ranges between $n$ to 10, the base events of these alerts are mapped to nodes in the tree. As shown in Figure 6, the HRCT receives the alerts from both the behavior-based and signature-based detection components. In the HRCT representation, each intrusion goal is represented by one node in the tree. The final goal of the intrusion may be disrupting some high-level system functionality, such as "Denial of service achieved against one or more of the OGTSS assets". This final step will be achieved through multiple small to moderate-sized steps. Successful execution of a step is looked upon as achieving an intermediate intrusion goal and captured as an HRCT node. The intrusion goals have dependency relationships between one another. The edges are used to model this kind of dependency.

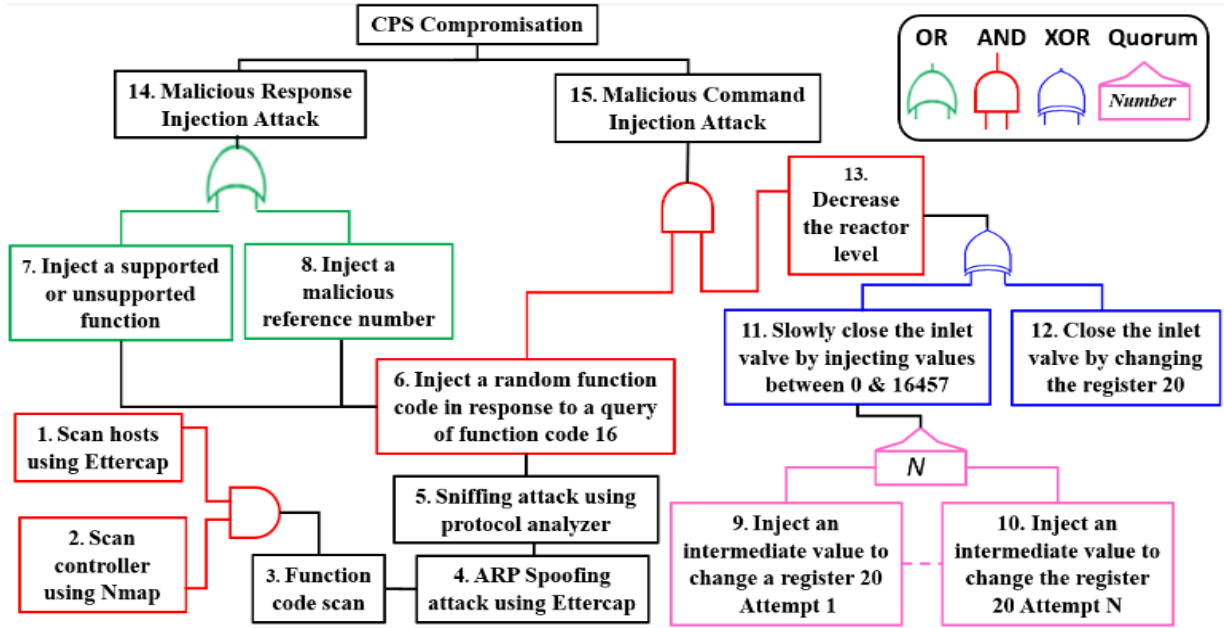

**Fig.6**: Real HRCT example

- **THE HRCT CONSTRUCTION ALGORITHM**

The following steps summarize the proposed method:

**Step 1: Construct the base-level and expanded vulnerability trees.** The top undesirable event is first postulated which represents a pivotal event for a particular failure scenario. The possible means (attacks) for this event to occur are systematically deduced by analyzing the collected alerts and logs. These attack paths can result in a failure event (the top event). Then, each situation (base events) that could cause an attack is added to the tree as a series of logical expressions. Thus, the intermediate failure events ("attacks") are connected to the top event and basic events with logic gates, the most common of which are "AND" gates, "OR" gates, and Quorum operation. In the HRCT, the AND gate is used when all the base-events connected by this gate must happen to launch an attack. The OR gate is used when any one of the base-events connected by this gate is sufficient for an intruder to launch an attack. The Quorum operation represents the minimum number of child nodes whose goals need to be achieved in order for the Quorum operation to be achieved. Conforming to the traditional definition of quorums in fault-tolerant systems (the minimum number of service replicas whose loss will affect the functionality of the service). As an example, in node 9 and 10 in the real HRCT of Fig. 6, one may think the Minimum Required Quorum (MRQ) is $N$ which denotes the minimum number of injections attempts an attacker should try before complete closure of an inlet valve. The overall risk value computed by the HRCT is obtained by the summation of all computed *CP* values.

**Step 2: Construct a damage analysis table based on each base event in the HRCT tree.** A list of all attack types and their corresponding base events that compromise our OGTSS testbed is created. Each of the base events is considered one at a time and a list of the financial asset costs affected by them is calculated. In this step, we build the financial damage analysis table with the corresponding impact dollar value for all base events and attacks that are described in Section 5. To build Table 7, we use for our calculation the information provided by the OGTSS domain experts and by studying literature on critical system attacks such as data in [43] and the British Columbia Institute of Technology Industrial Security Incident Database [44]. The latter database provides one of the most complete incident databases for cyber-attack records on critical systems. Using Table 7, we compute the financial damage of a particular attack and we involve the financial damage parameter in the risk assessment process.

Table 7: Example of Asset Values Corresponding to the HRCT Base Events

| Base Event | Main Affected Equipments | Cost of Damaged Equipments in $ | Loss of Control of the CPS Basic Functions ($) | Loss of Operator Salary in $ | Total Assets Cost |
|---|---|---|---|---|---|
| 6. Packet length is modified | MTU | 300 | 500 | 100 | $900 |
| 8. Inject one malformed TCP Packet | HMI and Controllers | 1 | 10 | 25 | $36 |
| 9. Inject N malformed TCP Packets |  | 600 | 400 | 500 | $1,500 |
| 10. Denial of Service (DoS) Attack | The connection between MTU and RTU | One Base Event * 100 | 1000 | 800 | $1,900 |

**Step 3: Compute the compromised probability at each node.** This step determines, based on the received alerts from the detectors, which of the HRCT goal nodes are likely to have been achieved. Each detector (behavior-based, or signature-based) provides confidence values for its alerts, termed alert reliability. If the detector does not provide an inbuilt confidence value with the alert, then the alert confidence value is set to one. When a detector flags an intrusion, the alerts are placed in the HRCT nodes with the corresponding intrusion event. The Compromised Probability (*CP*) of a node in the HRCT is a measure of the likelihood that the node has been achieved. It is computed based on the alert reliability corresponding to the node and the *CP* of its immediate children nodes. Mathematically, the *CP* of a node is given by Equation 4 as follows:

$$CP = \begin{cases} alert\ reliability, & Nodes\ that\ have\ no\ children \\ RI'(CP_i), & Nodes\ that\ have\ no\ detectors \\ RI(RI'(CP_i), alert\ reliability), & Otherwise \end{cases} \quad (4)$$

$$RI' = \begin{cases} Max(CP_i), & OR\ edge \\ Min(CP_i), & AND\ edge \\ \begin{cases} Mean\ (CP_i \mid CP_i \geq NT), & Quorum\ achieved \\ 0, & Quorum\ not\ achieved \end{cases} \end{cases}$$

Where,
- ($CP_i$) is the Compromised Probability of the $i^{th}$ child.
- $NT$ is a threshold per node. This threshold is computed for each node in the training phase and it represents the average alert severity at a certain node.

The intuition is that for an *OR* edge, the parent node can be achieved if any of its children nodes are achieved and therefore the likelihood is the maximum of that of all of its children. For an *AND* edge, all the children nodes have to be achieved and therefore the likelihood is as much as the least likely child node. For *Quorum* edges, if the quorum is not met, then the higher goal is not achieved, but if met, the likelihood of it being achieved only depends on the mean of that of its children nodes that have achieved the quorum. The Relative Important function *RI* allows various weights to be assigned to determine the relative importance placed to the alert or the position of the node in the HRCT. In our HRCT model, we update the *CP* using the alert reliability factor computed by the behavior-based or the signature-based detectors. The normalization of this factor is based on policy. For an aggressive policy, the maximum alert reliability in the alert queue is used; for a moderate policy, the maximum of a subset of alert reliability based on the most recent alerts is chosen; for a conservative policy, the alert reliability corresponding to the most recent alert is chosen. The alert reliability provided by a detector has to be moderated by the confidence of the detector. Our security framework has a mechanism to determine if a detector misses alarms and adjusts the detector's confidence accordingly. The framework finds that for a given node $n_i$, its children nodes, as well as parent nodes, are flagged but $n_i$ is not, then it anticipates probabilistically that the detectors have missed flagging the alert.

**Step 4: Compute the current risk related to the base event.** At each traversing node in the HRCT, the algorithm computes the risk corresponding to the base event of the current traversed node using Equation 5 and updates the overall risk value using Equation 6. The expanded tree now has information about threats, the severity and financial impact of these threats, and the vulnerability of the system to electronically launched attacks.

$$R_i = A_i * CP(e_i) \qquad (5)$$
$$R_{total} = \sum_{i=1}^{n} R_i \qquad (6)$$

– $A_i$: The sum of the asset values affected by the detected base event $e_i$. This value is computed in step 2 as shown in Table 7.

**Step 5: Repeat steps 2 to 4** to apply the approach to all nodes in HRCT till the top node is reached.

## 9. Risk Analysis Using the HRCT for the Command Injection Attack: Real-Time Case Study

We have performed the multistage attack scenarios described in section 5. We used reconnaissance attacks to gather information about the testbed (described in Section 3) and found useful parameters for building active attacks. In DoS attack, malformed TCP packet, and packet length attacks caused a denial of service. After performing this attack, the command-response cycle stopped, and MTU lost control over the RTU. Therefore, the legitimate user could not control the physical processes using MTU.

In response injection attack, injected random responses deceived MTU, interrupted the command-response cycle, and caused the control information leak. As a result, MTU stopped sending control commands to RTU, RTU sent several different responses that caused the control logic information leak, and the GUI (Graphical user interface) of HMI completely changed. In the command injection attack, we targeted the control valves that control the level and temperature of the reactor. The reactor temperature and level decrease or increase randomly by changing the position of the valves. For designing this attack, we first eavesdropped using protocol analyzers then figured out actual registers that control the actuators. Finally, we stopped the MTU to write those registers, then injected commands to change the values of those registers.

As a result, the reactor level and temperature increased or decreased according to the injected values. Figures 7, 8, and 9 show the impact of this attack on the physical processes. The experiment results in these figures are collected from the reactor simulation panel. The panel shows the real-time status of the reactor in terms of the reactor level and temperature that reflect the impact of the command injection attacks on the control valves. Fig. 7 shows the impact of the command injection attack on coolant valves CV-1 and CV-2. The attack causes the closure of CV-1 and the full opening of CV-2, this results in more coolant flow entered the reactor. Therefore, the reactor temperature dropped significantly. Fig. 8 and 9 show the disruption of the reactor level after the injection of malformed commands. In Fig. 8, the complete closure of CV-3 increased the reactor level from 50% (1.0) to 95% (1.9). Whereas in Fig. 9, the reactor level dropped from 50% (1.0) to 11% (0.22) because of the malfunction of CV-1.

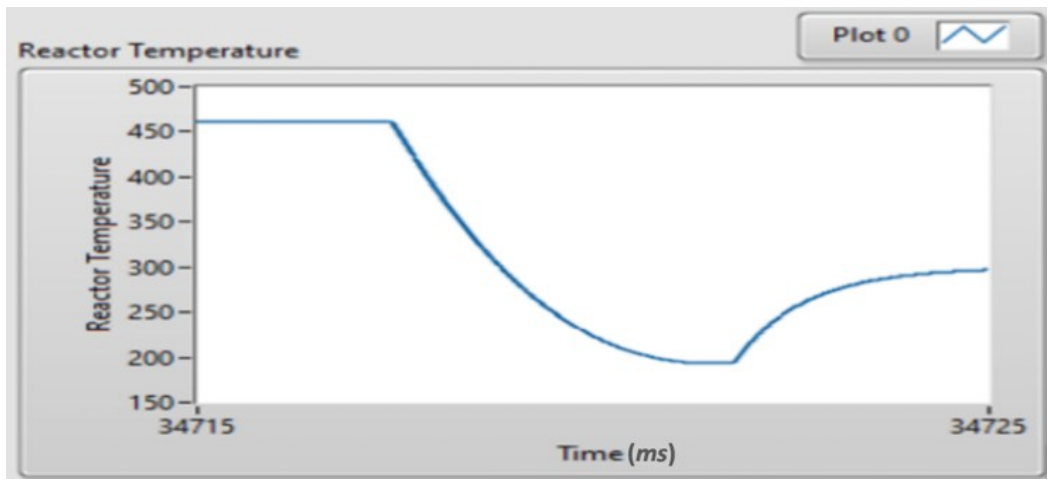

**Fig.7**: Injection attack on CV-1 and CV-2 valves

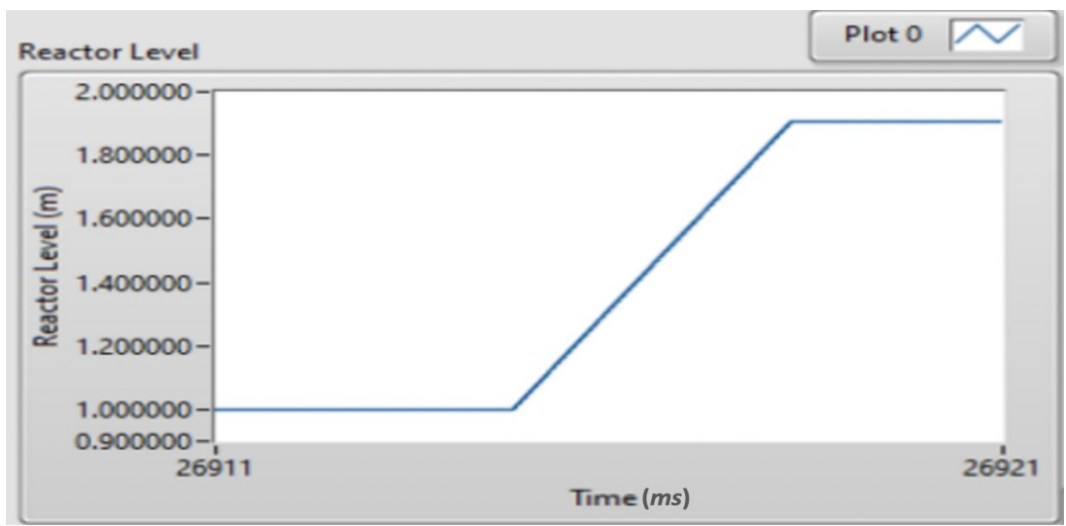

**Fig.8**: Injection attack on CV-3 valve

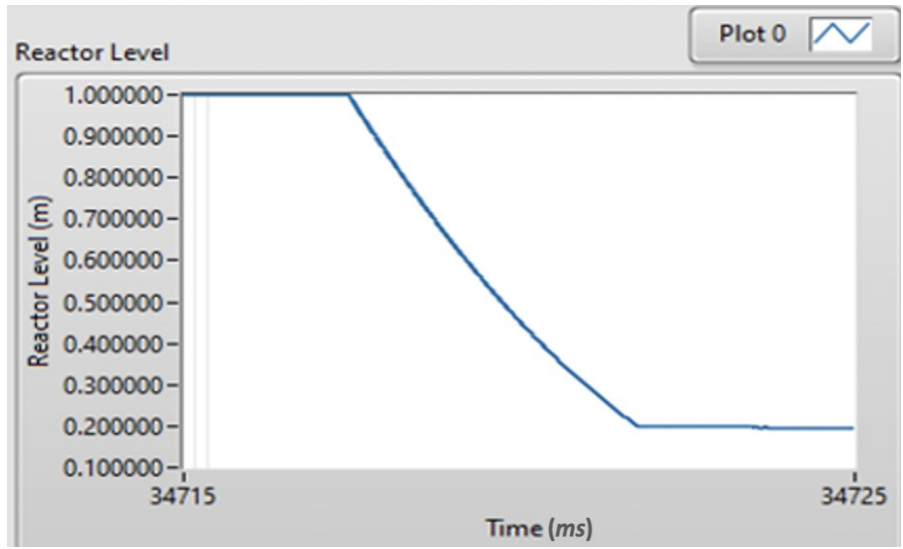

**Fig.9**: Injection attack on CV-1 valve

The following attack tree demonstrates an example of the HRCT model for the DoS attack tree, see Fig. 10.

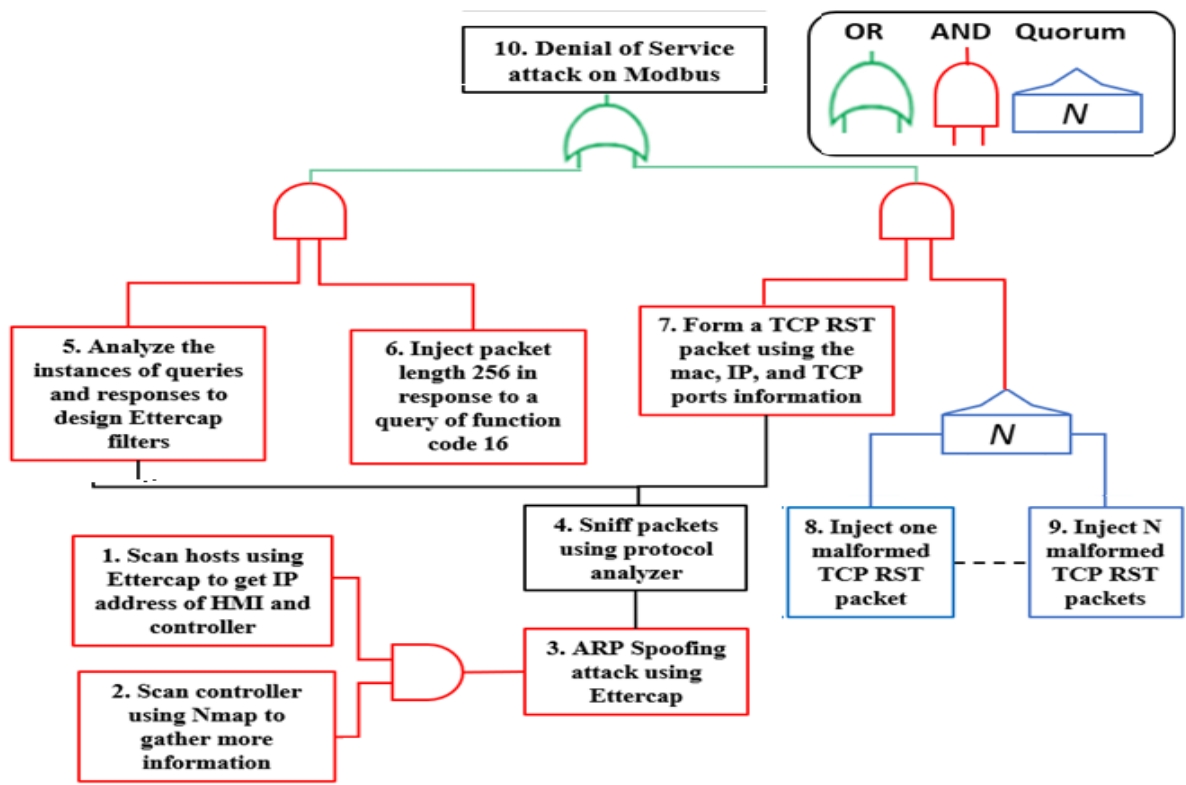

**Fig.10**: The DoS attack tree.

This tree shows the steps that an attacker took to disrupt the control process in the network. In this attack tree, we got alerts at nodes 1 and 2. Those alerts do not have an inbuilt confidence value. So, we are considering the alert reliability equal to 1 because the signature of the attacks is exactly matched. Furthermore, the *CP* value resulted from the *AND* edge of nodes 1 and 2 equals 1, *Min* (1, 1). At node 3, we did not get any alerts after performing ARP spoofing attack. Therefore, according to

Equation 4, the CP value at node 4 is 1. At node 6, after we injected the packet of length 256KB in a response to a query of function code 16, we got the alert of incorrect packet length. Now, we are considering *CP* value at node 6 equal to 1 because the signature of the attack is exactly matched. The *CP* value resulted from the *AND* edge of nodes 5 and 6 equals *1, Min(1, 1 )*. At nodes 8 and 9, we are using a minimum type of Quorum gate. We were injecting 10000 packets at a time, after inserting 7763 packets, the TCP connection stopped. Thus, the *CP* value resulted from the *Quorum* gate is 0.7 (7763/10000). As a result, the *CP* value resulted from the *AND* edge of nodes 7, 8, and 9 equal to 0.7, *Min* (1, 0.7). Finally, the CP value resulted from the OR gate is *1, Max (1,0.7).* To compute the total risk, we calculate the CP values using Equations 4, 5, and 6. As shown in table 7, we estimate the asset values affected by the detected events at nodes 6, 8, and 9. These assets include the MTU, RTU, and the network connection between them. Some aspects are considered when calculating asset values. These aspects include the cost of a damaged component, financial loss due to system downtime, and repair cost. Suppose that the estimated asset values affected are $900 at node 6 and $1500 at nodes 8 and 9. Considering the calculated CP values (i.e., 1 and 0.7), therefore using Equations 5 and 6, the *R_total* =1950.

To compare the accuracy of our HRCT model with the Suricata [37] intrusion response system, we use the evaluation models, (a) the Mean Absolute Percentage Error (MAPE) model [38], (b) the Maxion-Townsend Cost function [39], and a VEA-bility Metric [40].

*(a)    Compare the accuracy of our HRCT with the Suricata using the MAPE Model.*

In our experiments, The Suricata IPS is deployed in our testbed described in Section 3. We run three categories of attack scenarios described in Section 5.

Table 8 denotes the MAPE equation whereas Table 9 denotes, the MAPE error rate for the HRCT model is lower than Suricata IPS for the DOS, Response Injection, and Command Injection attacks by 59.80%, 73.72%, and 66.96% respectively.

TABLE 8: MAPE EQUATION

| Metric | Calculation |
|---|---|
| MAPE (Mean Absolute Percentage Error) | $\frac{1}{n}\sum_{i=1}^{n}\frac{|a_i - c_i|}{a_i}$ where $a_i$ and $c_i$ are the actual and computed values respectively |

TABLE 9: The MAPE Error rate for the HRCT and Suricata IPS

| Attack Name | Accuracy Metric | HRCT Risk Accuracy | Suricata IPS |
|---|---|---|---|
| DOS | MAPE | 0.164 | 0.408 |
| Response Inection | MAPE | 0.103 | 0.392 |
| Command Injection | MAPE | 0.150 | 0.454 |

*(b)    Compare the accuracy of our HRCT model with the Suricata using the Maxion-Townsend Cost Function*

The Maxion-Townsend cost function defines the detection cost of the ROC curve in Figure 11 according to Equation 7.

*Cost = 6* FalsePositive Rate + (100-HitRatio)*          (7)

Figure 11 and Table 10 show that the HRCT has a lower impact on the overall detection accuracy than the Suricata IPS. The HRCT model has a much lower Maxion-Townsend cost comparing to the Suricata IPS.

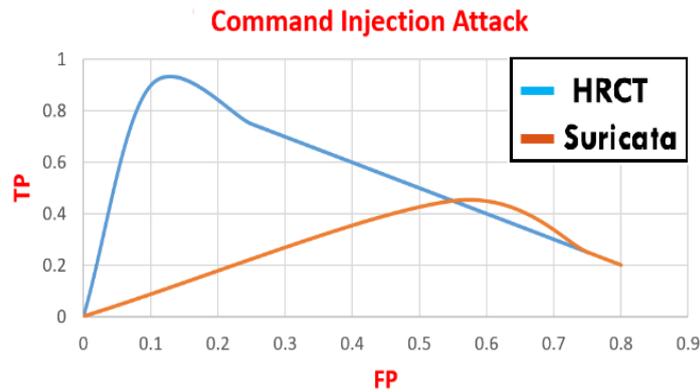

Fig. 11: ROC curve of command injection attack

TABLE 10: MAXION-TOWNSEND COST for the HRCT and Suricata IPS

| Detection Model | False Positive Rate(%) | Hit Ratio(%) | Maxion-Townsend Cost |
|---|---|---|---|
| **HRCT** | 13 | 90 | 70.6 |
| **Suricata IPS** | 55 | 45 | 385 |

Fig. 12 shows the difference between risk calculations of the HRCT and Suricata IPS. The risk computed by HRCT is higher than that computed by Suricata due to the high True Positive (*TP*) rates and low False Positive (*FP*) rates of the HRCT.

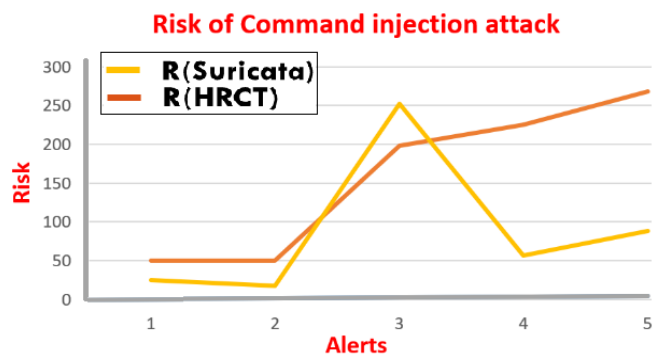

Fig. 12: Risk comparison of command injection attack

*(c)     Compare the accuracy of the HRCT model with the Suricata Using the VEA-bility Metric.*

The underlying idea behind the VEA-bility metric is that the security of a network is influenced by many factors, including the severity of existing vulnerabilities, distribution of services, connectivity of hosts, and possible attack paths. These factors are modeled into three network dimensions: Vulnerability, Exploitability, and Attackability. The overall VEA-bility score, a numeric value in the range [0,10], is a function of these three dimensions, where lower value implies better security. The VEA-bility metric uses data from three sources: the OGTSS network topology, attack graphs, and scores as assigned by the Common Vulnerability Scoring System (CVSS) [41]. To adjust the VEA-bility metric to validate the risk evaluation of the HRCT and Suricata IPS, we modify this metric by replacing the asset Attackability factor by the Compromised Probability (*CP*) value at Equation 4 for each event $e_i$. We let each vulnerability *v*, that corresponds to an *event* $e_i$, have an impact score, exploitability score, and temporal score as defined by the CVSS [41]. An impact and exploitability subscores are automatically generated for each common vulnerabilities identifier based on its CVE name defined by the CVSS, whereas the temporal score requires user input. We then define the severity, *S*, of a vulnerability to be the average of the impact and temporal scores, Eq. (8):

$$S(v) = (Impact\ Score(v) + Temporal\ Score(v)) / 2 \qquad (8)$$

The Vulnerability score (*V*) of an OGTSS computing asset, e.g., PMU, PDC, etc is an exponential average of the severity scores of the vulnerabilities on the OGTSS asset, or 10, whichever is lower. The asset Exploitability score (*E*) is the exponential average of the exploitability score for all asset vulnerabilities multiplied by the ratio of network services on the asset. The asset Attackability score (*A*) refers to the toral CP values for all vulnerabilities at a certain asset. The Attackability score is multiplied by a factor of 10 to produce a number in the range [0,10], ensuring that all dimensions have the same range. For an asset, *a*, let *v* be an asset vulnerability. We then define the three asset dimensions as shown in equations (9), (10), and (11):

$$V(a) = min(10, \ln \sum e^{S(v)}) \qquad (9)$$

$$E(a) = (min(10, \ln \sum e^{Exploitability\ Score(v)}))\ (\#\ services\ on\ a)/(\#\ network\ services) \qquad (10)$$

$$A(t) = (10) * \sum_{i=1}^{n} a_{CP(e_i)} \qquad (11)$$

The overall VEA-bility equation for *a* then becomes as in Eq. (12).

$$VEA\text{-}bility_a = 10 - ((V+E+A)_a / 3) \qquad (12)$$

To test the performance of the proposed VEA-bility metric for both the HRCT and Suricata IPS, we developed an extensive set of scenarios described in section 5 and used the vulnerabilities observed by Nessus [42] scan results after running the attacks scenarios. Figure 13 shows the overall average VEA-bility scores observed in our experiments for the OGTSS assets. A higher score indicates a more secure configuration, which we call more "VEA-ble". According to Figure 13, HRCT has higher VEA-ble scores than the Suricata IPS, on average HRCT is 17.02% more VEA-ble than the Suricata IPS.

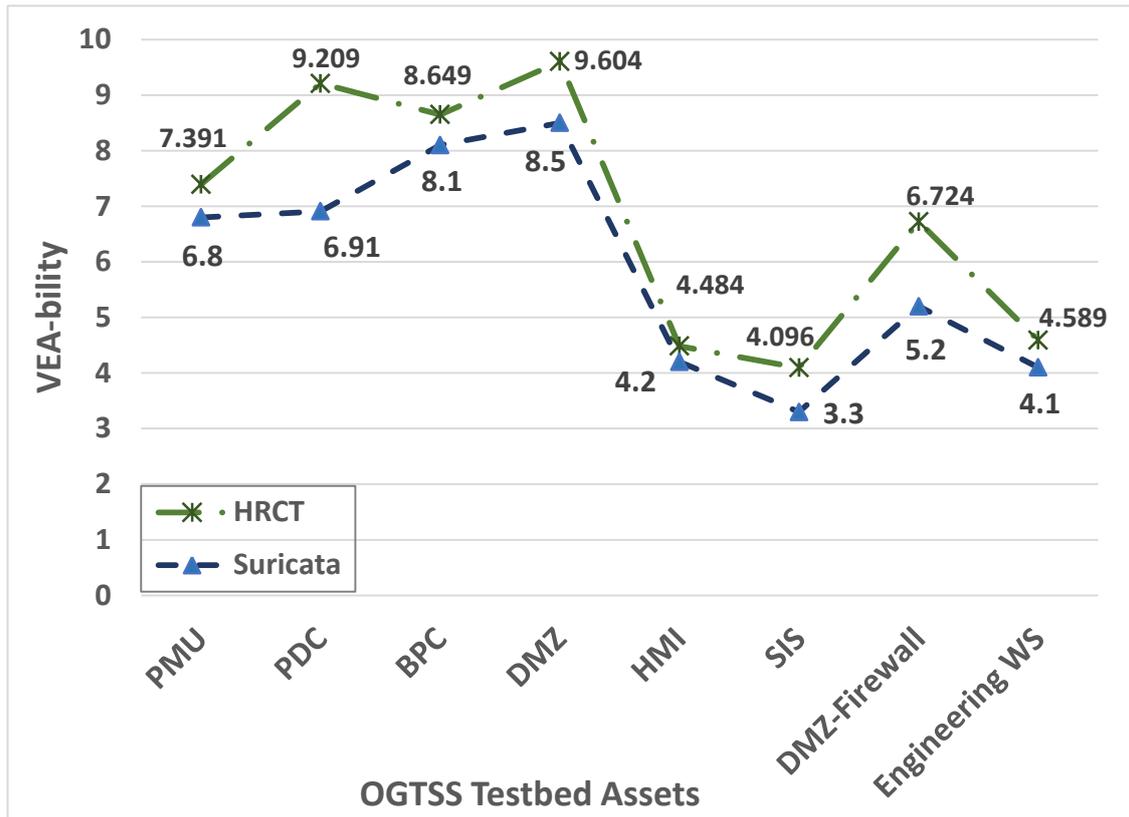

Fig. 13: The VEA-bility metric values of the HRCT and the Suricata IPS

## 10. CONCLUSION AND FUTURE WORK

The OGTSS has become one of the vital cyber-physical systems. However, there are increasing security assessment requirements for the OGTSS, specifically to achieve compliance requirements for regulatory agencies. This paper introduces two main contributions, (i) a Data Quantization and State Compression Approach (DQSCA) that improves the classification accuracy and speeds up the detection algorithm. The experiment results depict that DQSCA reduces the detection accuracy a little bit (2.45% and 4.45%) while reduces the detection time much (27.74% and 42.06%) for the Turnipseed's and Gao's datasets respectively. (ii) A quantitative risk assessment model is a probabilistic model that is built on the fact that complex or multi-stage attacks are a sequence of elementary attacks where a threat agent acquires the privileges to implement each attack through the previous threats in the chain. The experiment results depict that HRCT successfully analyses the risk in the OGTSS network with a low error rate. We evaluated the model using the MAPE models for three of the attacks covered by the HRCT namely, the DoS, response injection, command injection. It achieves the lowest MAPE error rate of 0.103 for response injection attack. The highest error rate of the HRCT is still acceptable in the cybersecurity of the OGTSS domain. The HRCT has, on average, higher VEA-ble scores than the Suricata IPS by 9.53%.

In future work, we will also evaluate the computational performance of the risk assessment model in a large scale OGTSS. We will also integrate the HRCT with an autonomous intrusion response system that considers the risk values to deploy countermeasures against ongoing attacks on the OGTSS.

**Supplementary Materials:** NA.

**Author Contributions:** Sole Author.

**Funding:** This research was generously supported in part by the SUNY Polytechnic Institute Research Seed Grant Program.

**Conflicts of Interest:** "The authors declare no conflict of interest."

**References**

1. Figueroa-Lorenzo S, Añorga J, Arrizabalaga S. A Role-Based Access Control Model in Modbus SCADA Systems. A Centralized Model Approach. Sensors (Basel). 2019;19(20):4455. Published 2019 Oct 14. doi:10.3390/s19204455.
2. G Yadav, K Paul, "Architecture and Security of SCADA Systems: A Review", arXiv preprint arXiv:2001.02925, 2020.
3. Knapp, Eric D.: Industrial Network Security: securing critical infrastructure networks for smart grid, SCADA and other industrial control system, pp.30–152. Elsevier Inc, Waltham (2014)
4. Li Yang, Xiedong Cao, Xinyu Geng, "A novel intelligent assessment method for SCADA information security risk based on causality analysis", Journal of Cluster Computing, DOI 10.1007/s10586-017-1315-4.
5. Wang J.; Rong, L.; "Cascade-based attack vulnerability on the US power grid," Safety Sci., vol. 47, no. 10, pp. 1332– 1336, Dec. 2009.
6. Hisham A. Kholidy, Abdelkarim Erradi, "A Cost-Aware Model for Risk Mitigation in Cloud Computing Systems", the 12th ACS/IEEE International Conference on Computer Systems and Applications (AICCSA), Marrakech, Morocco, November 2015.
7. Hisham A. Kholidy, Abdelkarim Erradi, "VHDRA: A Vertical and Horizontal Dataset Reduction Approach for Cyber-Physical Power-Aware Intrusion Detection Systems", SECURITY AND COMMUNICATION NETWORKS Journal (ISI IF: 1.376), March 7, 2019. vol. 2019, Article ID 6816943, 15 pages. https://doi.org/10.1155/2019/6816943.-
8. Hisham A. Kholidy, Ali Tekeoglu, Stefano Lannucci, Shamik Sengupta, Qian Chen, Sherif Abdelwahed, John Hamilton, "Attacks Detection in SCADA Systems Using an Improved Non-Nested Generalized Exemplars Algorithm", the 12th IEEE International Conference on Computer Engineering and Systems (ICCES 2017), December 19-20, 2017.
9. Farahmand, F., Navathe, S. B., Sharp, G. P., & Enslow, P. H. (2005). A management perspective on risk of security threats to information systems. Information Technology and Management, 6(2–3), 203–225.
10. K. L. Deng et al., "Study on a Quantitative Risk Assessment Method of Primary Equipments Based on Power Grid Safety Analysis", Applied Mechanics and Materials, Vols. 541-542, pp. 863-868, 2014
11. National Infrastructure Protection Plan, https://www.dhs.gov/xlibrary/assets/NIPP_RiskMgmt.pdf, January 2018
12. Yulia Cherdantseva, Pete Burnap, Andrew Blyth, Peter Eden, Kevin Jones, Hugh Soulsby, and Kristan Stoddart, "A review of cyber security risk assessment methods for SCADA systems", Computers & Security Volume 56, February 2016.
13. M. Nasser, R. Ahmad, W. Yassin et al., "Cyber-security incidents: a review cases in cyber-physical systems," International Journal of Advanced Computer Science and Applications, vol. 9, no. 1, 2018.
14. P. S. Woo and B. H. Kim, "A study on quantitative methodology to assess cyber security risk of SCADA systems," Advanced Materials Research, vol. 960-961, pp. 1602–1611, 2014.
15. Hisham A. Kholidy, Abdelkarim Erradi, "VHDRA: A Vertical and Horizontal Intelligent Dataset Reduction Approach for Cyber-Physical Power Aware Intrusion Detection Systems", Security and Communication Networks, vol. 2019, Article ID 6816943, 15 pages, 2019. https://doi.org/10.1155/2019/6816943
16. Kholidy, H. A. (2020), "Autonomous mitigation of cyber risks in the Cyber–Physical Systems", doi:10.1016/j.future.2020.09.002, Future Generation Computer Systems,Volume 115, 2021, Pages 171-187, ISSN 0167-739X, https://doi.org/10.1016/j.future.2020.09.002. (http://www.sciencedirect.com/science/article/pii/S0167739X19320680).
17. Kholidy, H.A., "Detecting impersonation attacks in cloud computing environments using a centric user profiling approach", Future Generation Computer Systems, Volume 115, issue 17, December 13, 2020, Pages 171-187, ISSN 0167-739X, https://doi.org/10.1016/j.future.2020.12 , https://www.sciencedirect.com/science/article/abs/pii/S0167739X20330715
18. Hisham A. Kholidy, "Detecting impersonation attacks in cloud computing environments using a centric user profiling approach", Future Generation Computer Systems, Volume 117, 2021, Pages 299-320, ISSN 0167-739X, https://doi.org/10.1016/j.future.2020.12.009. (http://www.sciencedirect.com/science/article/pii/S0167739X20330715)

19. Kholidy, Hisham A.: 'Correlation-based sequence alignment models for detecting masquerades in cloud computing', IET Information Security, 2020, 14, (1), p. 39-50, DOI: 10.1049/iet-ifs.2019.0409. IET Digital Library, https://digital-library.theiet.org/content/journals/10.1049/iet-ifs.2019.0409
20. H. A. Kholidy, "Towards A Scalable Symmetric Key Cryptographic Scheme: Performance Evaluation and Security Analysis," 2019 2nd International Conference on Computer Applications & Information Security (IC-CAIS), Riyadh, Saudi Arabia, 2019, pp. 1-6, doi: 10.1109/CAIS.2019.8769494.
21. Kholidy, H.A., Fabrizio Baiardi, Salim Hariri: 'DDSGA: A Data-Driven Semi-Global Alignment Approach for Detecting Masquerade Attacks'. The IEEE Transaction on Dependable and Secure Computing, 10.1109/TDSC.2014.2327966, pp:164–178, June 2015.
22. Kholidy, H.A., Fabrizio Baiardi, "CIDD: A Cloud Intrusion Detection Dataset For Cloud Computing and Masquerade Attacks ", in the 9th Int. Conf. on Information Technology: New Generations ITNG 2012, April 16-18, Las Vegas, Nevada, USA. http://www.di.unipi.it/~hkholidy/projects/cidd/
23. Kholidy, H.A., Fabrizio Baiardi, "CIDS: A framework for Intrusion Detection in Cloud Systems", in the 9th Int. Conf. on Information Technology: New Generations ITNG 2012, April 16-18, Las Vegas, Nevada, USA. http://www.di.unipi.it/~hkholidy/projects/cids/
24. Kholidy, H.A., Baiardi, F., Hariri, S., et al.: 'A hierarchical cloud intrusion detection system: design and evaluation', Int. J. Cloud Comput., Serv. Archit. (IJCCSA), 2012, 2, pp. 1–24.
25. Hisham A. Kholidy, Abdelkarim Erradi, Sherif Abdelwahed, Abdulrahman Azab, "A Finite State Hidden Markov Model for Predicting Multistage Attacks in Cloud Systems", in the 12th IEEE International Conference on Dependable, Autonomic and Secure Computing (DASC), Dalian, China, August 2014.
26. Kholidy, H.A., Abdelkarim Erradi, Sherif Abdelwahed, Fabrizio Baiardi, "A risk mitigation approach for autonomous cloud intrusion response system", in Journal of Computing, Springer, DOI: 10.1007/s00607-016-0495-8, June 2016.
27. Kholidy, H.A., Abdelkarim Erradi, "A Cost-Aware Model for Risk Mitigation in Cloud Computing SystemsSuccessful accepted in 12th ACS/IEEE International Conference on Computer Systems and Applications (AICCSA), Marrakech, Morocco, November, 2015.
28. Kholidy, H.A., Ali Tekeoglu, Stefano Lannucci, Shamik Sengupta, Qian Chen, Sherif Abdelwahed, John Hamilton, "Attacks Detection in SCADA Systems Using an Improved Non-Nested Generalized Exemplars Algorithm", the 12th IEEE International Conference on Computer Engineering and Systems (ICCES 2017), December 19-20, 2017.
29. Qian Chen, Kholidy, H.A., Sherif Abdelwahed, John Hamilton, "Towards Realizing a Distributed Event and Intrusion Detection System", the International Conference on Future Network Systems and Security (FNSS 2017), Gainesville, Florida, USA, 31 August 2017. Conference publisher: Springer. "Industrial control system (ics) cyberattack datasets, http://www.ece.uah.edu/~thm0009/icsdatasets/ PowerSystem_Dataset_README.pdf
30. Kholidy, H.A. Multi-Layer Attack Graph Analysis in the 5G Edge Network Using a Dynamic Hexagonal Fuzzy Method. Sensors 2022, 22, 9. https://doi.org/10.3390/s22010009
31. Haque, N.; Rahman, M.; Chen, D.; Kholidy, H. BIoTA: Control-Aware Attack Analytics for Building Internet of Things. In Proceedings of the 18th IEEE International Conference on Sensing, Communication and Networking (SECON), Rome, Italy, 6–9 July 2021.
32. Morris, T., Vaughn, R., Dandass, Y., "A Retrofit Network Intrusion Detection System for MODBUS RTU and ASCII Industrial Control Systems", Proceedings of the 45th IEEE Hawaii International Conference on System Sciences (HICSS – 45). January 4-7, 2012. Grand Wailea, Maui.
33. H. Debar, D. Curry, "The Intrusion Detection Message Exchange Format (IDMEF)", rfc4765, March 2007.
34. Daniela Z., Lavinia P., Viorel N. and Flavia Z., "Evolutionary Pruning of Non-Nested Generalized Exemplars", 6th IEEE Int. Symposium on Applied Computational Intelligence and Informatics, May 19–21, 2011, Timişoara, Romania.
35. Ian P Turnipseed, "A new SCADA dataset for intrusion detection system research", 2015, Master Thesis, Mississippi State University http://sun.library.msstate.edu/ETD-db/theses/available/etd-06292015-115535/
36. T. Morris; W. Gao, Industrial Control System Network Traffic Data Sets to Facilitate Intrusion Detection System Research, Mississippi State University.
37. Suricata IPS, https://suricata-ids.org/
38. MAPE Arnaud De Myttenaere, Boris Golden, Bénédicte Le Grand, Fabrice Rossi, "Mean absolute percentage error for regression models", Neurocomputing 2016 , DOI. 10.1016/j.neucom.2015.12.114, volume 192, ISSN= 0925-2312, pp. 38–48.
39. R. A. Maxion and T. N. Townsend, "Masquerade detection using truncated command lines," in Proc. Int. Conf. Dependable Syst. Netw., Washington, DC, USA, Jun. 2002, pp. 219–228.

40. Tupper M, Zincir-Heywood A (2008) VEA-bility security metric: a network security analysis tool. In: Proc IEEE Third Int'l Conf. Availability, Reliability and Security.
41. A Complete Guide to the Common Vulnerability Scoring System (CVSS): http://www.first.org/cvss/v1/guide.html.
42. Nessus Vulnerability Scanner: http://www.nessus.org.
43. Rakaczky, E. (2005). Building a security business case. Process control systems forum, October 25–27, Chicago,Illinois,_www.pcsforum.org/events/2005/fall/pdf/Building%20a%20Security%20Business%20Case2a.pdf, Jan 5, 2007.
44. The Industrial Security Incident Database, http://www.risidata.com/Database.

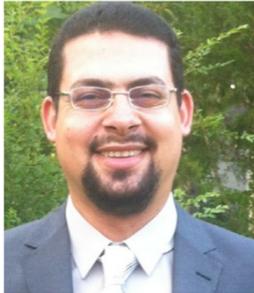

Hesham Abdelazim Ismail Mohamed (Hisham A. Kholidy) received his Ph.D. in Computer Science in a joint Ph.D. program between the University of Pisa in Italy and the University of Arizona in the USA. He works as an assistant professor at the Department of Networks and Computer Security (NCS), College of Engineering, State University of New York (SUNY) Polytechnic Institute. Before that, he worked as a postdoctoral associate at the University of Nevada, Reno, and Mississippi State University. During his Ph.D., he worked as an associate researcher at the NSF Cloud and Autonomic Computing Center, Electrical and Computer Engineering Dept. at the University of Arizona. He holds several patents in Cybersecurity published by the United States Patent and Trademark Office (USPTO), and he published more than 50 papers on high-quality journals and conferences. He received several research awards such as the SUNY Poly dean's excellence research award, AFRL VFRA award, and SUNY Seed Research Award. He participated as PI, Co-PI, and senior personnel in several research projects and his research interests include Cybersecurity and SCADA systems security, 5G systems security, Cloud Computing and high-performance systems, service composition, and big data analytics.